\documentclass{jfm}

\usepackage{graphicx}
\usepackage{float} %figure positioning
\usepackage{newtxtext}
\usepackage{newtxmath}
\usepackage{natbib}
\usepackage{hyperref}
\usepackage{xcolor}
\usepackage{longtable}
\usepackage{bm}
\hypersetup{
    colorlinks = true,
    urlcolor   = blue,
    citecolor  = green!50!black,
}

\newcommand{\RomanNumeralCaps}[1]
\newcommand{\jump}[1]{\left[\!\left[ #1 \right]\!\right]}

\title{Surfactant-induced Marangoni stresses and surface viscosity in Faraday instability
}

\author{B. Wang \and C. R. Constante-Amores%
\thanks{\texttt{crconsta@illinois.edu}}}

\affiliation{Department of Mechanical Science and Engineering,
University of Illinois Urbana-Champaign, USA}

\begin{document}
\maketitle

\begin{abstract}
We investigate the influence of insoluble-surfactant elasticity and Boussinesq-Scriven surface viscosity on the parametric instability of an interface separating two viscous fluids. The two-fluid Floquet formulation of Kumar \& Tuckerman (\textit{J. Fluid Mech.}, vol. 279, 1994, pp. 49–68) is extended to include surfactant transport, Marangoni stresses and surface viscosity. For the parameter range considered, both mechanisms increase the onset acceleration relative to the clean interface state, but through distinct responses: Marangoni elasticity produces a non-monotonic threshold, whereas surface viscosity yields a monotonic increase towards saturation.
Analysis of the corresponding unforced gravity-capillary modes shows that both mechanisms modify modal damping and natural frequency. Wavenumber selection reduces the detuning from subharmonic resonance by approximately an order of magnitude relative to that evaluated at the clean interface critical wavenumber. 
Along the selected branches, variations in the onset acceleration closely follow those in modal damping, with residual detuning playing a secondary role. Despite their distinct interfacial stress mechanisms, both Marangoni elasticity and surface viscosity suppress tangential interfacial motion in the strong-rheology limit, approaching a common tangentially immobilized condition.

\end{abstract}

\section{Introduction}

Parametrically excited interfacial waves, commonly known as Faraday waves, provide a canonical system for studying parametric instability and pattern formation \citep{faraday1831xvii,benjamin1954stability,miles1990parametrically}.
For two viscous fluids, \citet{kumar1994parametric} established the full Floquet formulation required to determine the onset acceleration and selected wavenumber, showing that viscous effects cannot in general be represented by a simply damped Mathieu equation.

In many realistic situations, fluid interfaces are intentionally or naturally contaminated with surface-active agents (surfactants) that modify interfacial dynamics by creating gradients in surface tension (i.e., Marangoni stresses) and introducing additional interfacial stresses (surface rheology) \citep{bird2002transport,brenner2013interfacial}. 
\cite{kumar2002parametrically} analysed the linear stability of Faraday waves in a liquid layer with an insoluble surfactant under free-surface assumption. 
The coupling between mean surfactant concentration and velocity perturbations is absent in the linearized equations, requiring an ad hoc relation between free-surface deformation and surfactant concentration. They  showed that surfactants can either increase or decrease the critical vibration amplitude, depending on their phase relative to the interface. Next,  \citet{kumar2004faraday} incorporated surfactant advection in the linearized formulation, eliminating the need for imposed coupling.
This revealed that the critical forcing amplitude can exhibit non-monotonic dependence on the Marangoni number due to the dynamically determined phase relationship between surfactant concentration and surface deformation.

Experiments also demonstrate that the influence of surfactant is closely connected to the dynamics of the underlying gravity–capillary mode. \citet{HENDERSON_1998} measured the damping rates and natural frequencies of surfactant-covered Faraday waves and showed that both quantities vary strongly with monolayer properties. Numerical calculations by \citet{ubal2005elastic} subsequently confirmed that surfactant elasticity can substantially modify the forcing required for onset and connected this response to the evolving interfacial concentration and flow fields.
All of these studies neglect surface rheological effects.
% Furthermore, numerical simulations by \citet{ubal2005elastic} examined the nonlinear evolution of surfactant-laden Faraday waves. They showed that 
% % the presence of 
% surfactant increases the forcing amplitude required to generate standing waves. They further demonstrated that this stabilizing effect is associated with Marangoni stresses, which generate a temporal phase shift between the surfactant concentration and the free-surface deformation.  Using reduced models based on lubrication theory, 
% \citet{kumar2002instability} showed that while a clean film is linearly stable, the presence of surfactant can induce instability through Marangoni stresses. All of these studies neglect surface rheological effects.
Surface viscosity introduces a qualitatively different interfacial response.  For a free-surface, \citet{ubal2005influence} found a sigmoidal increase of Faraday onset with surface viscosity, and \citet{giavedoni2007onset} considered simultaneous elastic and viscous surface stresses. More recently, \citet{strickland2024role}
examined Marangoni and Boussinesq contributions asymptotically and emphasized their connection to wave-energy dissipation.

% Numerical and linear-stability studies of surfactant-covered free surfaces have shown that Boussinesq-Scriven stresses can substantially alter Faraday-wave onset and may produce a sigmoidal dependence of the threshold on surface viscosity \citep{ubal2005influence,giavedoni2007onset,strickland2024role}. More recently, fully two-phase simulations have examined nonlinear surfactant-laden Faraday waves \citep{panda2025marangoni}.

To our knowledge, a Floquet formulation for the interface between two viscous fluids that combines insoluble-surfactant transport with Boussinesq–Scriven surface viscosity has not been used to separate these effects systematically. Here we extend the framework of \citet{kumar1994parametric} to determine how Marangoni stresses and surface viscosity modify the onset acceleration and selected wavenumber. To relate their distinct effects on the instability threshold to changes in the damping and frequency of the corresponding unforced gravity--capillary mode. We then use the critical Floquet eigenvectors to characterize the limiting interfacial response at strong elasticity and surface viscosity. 
% and to establish whether the two mechanisms approach a common hydrodynamic limit under strong interfacial resistance. 
% We present results in air-water systems properties at varying viscosity ratios. 
Section 2 presents the LSA, Section 3 summarizes the results, and final remarks are
presented 
in Section 4.

\section{Linear stability analysis}

\subsection{Governing equations}

We consider two immiscible incompressible fluids of depths $h_1$ and $h_2$, separated by an initially flat interface at $z=0$. The lower and upper fluids are denoted by $j=1$ and $j=2$, respectively, with density $\rho_j$ and dynamic viscosity $\eta_j$. Thus, $\rho_1$ and $\eta_1$ denote the properties of the lower fluid, whereas $\rho_2$ and $\eta_2$ denote those of the upper fluid, with $\rho_1>\rho_2$. In a frame oscillating vertically with the imposed vertical acceleration $G(t)=g_0+a\cos\Omega t$, the bulk equations are
\begin{align}
\rho_j (\partial_t + \boldsymbol{u}_j \cdot \nabla)\boldsymbol{u}_j 
&= -\nabla p_j + \eta_j \nabla^2 \boldsymbol{u}_j - \rho_j G(t)\boldsymbol{e}_z, \label{eq:NS}\\
\nabla \cdot \boldsymbol{u}_j &= 0,
\label{eq:contin}
\end{align}
\noindent where $\boldsymbol{u}_j$ and $p_j$ denote the velocity and pressure fields, respectively.

The interface carries an insoluble surfactant of dimensionless concentration $\Gamma=\hat{\Gamma}/\Gamma_\infty$, governed by
\begin{equation}
    \partial_t \Gamma + \nabla_s\cdot(\Gamma \bm u_s)+(\bm u_s \cdot \boldsymbol{n})(\nabla_s \cdot \boldsymbol{n})\Gamma= D_s \nabla_s^2 \Gamma,
    \label{eq:surf_transport_exact}
    \end{equation}
where $\boldsymbol{u}_s$ is the interfacial velocity; $D_s$ is the surface diffusivity; $\nabla_s$ denotes the surface gradient operator and $\Gamma_\infty$ is the maximum packing surfactant concentration. Surface tension follows the Langmuir equation, i.e.,
% \begin{equation}
$\sigma(\Gamma)=\sigma_0\left[1+\beta_s\ln(1-\Gamma)\right]$,
\label{Langmuir_model}
% \end{equation}
where $\beta_s={RT\Gamma_\infty}/{\sigma_0}$ is the dimensionless surface elasticity parameter and $\sigma_0$ is the surfactant-free surface tension. 
The base-state surface tension for
an initial uniform
% base state with a initial
surfactant concentration $\Gamma_0$ is therefore $\sigma_b\equiv\sigma(\Gamma_0)$.
% \begin{equation}
%\sigma_b\equiv\sigma(\Gamma_0)=\sigma_0\left[1+\beta_s\ln(1-\Gamma_0)\right].
% \label{eq:sigma_base}
% \end{equation}

We denote jumps across the interface by $\jump{f}=f_2-f_1$ where $f$ may be a scalar, vector, or tensor field evaluated on the two sides of the interface. The continuity of velocity,  the kinematic condition, and the interfacial stress balance are written as

\begin{align}
\jump{\boldsymbol{u}} = 0,\\
[\partial_t+(\boldsymbol{u}\cdot\nabla)] \zeta = w|_{z=\zeta}, \\
 \jump{\mathbf{T}} \cdot \boldsymbol{n} = \nabla_s \cdot \mathbf{T}^S,
\end{align}
\noindent where $\boldsymbol{n}$ is the unit normal to the interface and the stress tensor in each phase is
\begin{equation}
    \mathbf{T}_{j,lm} = -(p_j-\rho_jG(t)z)\delta_{lm}  + \eta_j(\partial_l\bm u_{j,m} + \partial_m \bm u_{j,l}).
    \label{eq:stress}
\end{equation}
Here, the indices $l,m$ refer to components $x,y , z.$ 
% and $j$ to the fluid layer.
Surface rheological effects are described by the Boussinesq-Scriven constitutive equation  \citep{scriven1960dynamics},

\begin{equation}
        \mathbf{T}^S=\sigma \mathbf{I}^S+\left(\kappa^S - \eta^S\right)\nabla_s\cdot\boldsymbol{u}_s \mathbf{I}^S+\eta^S
        \left[
        \nabla_s \boldsymbol{u}_s \cdot \mathbf{I}^S+
        \mathbf{I}^S \cdot \left(\nabla_s \boldsymbol{u}_s\right)^T
        \right]
    \label{eq:boussinesq_scriven}
    \end{equation}
 where superscript $S$ denotes interfacial constitutive quantities associated with the Boussinesq-Scriven model and $\kappa^S$ and $\eta^S$ are the dilatational and shear surface viscosities, respectively,
%
% The interfacial velocity $\mathbf{u}^S$ can be decomposed into tangential and normal components as
% \begin{equation}
% \mathbf{u}^S = \mathbf{u}_s+ (\mathbf{u}^S \cdot \mathbf{n})\mathbf{n},
% \end{equation}
% where $\mathbf{u}_s=\mathbf{u}^S \cdot \mathbf{I}^S$ is the interfacial tangential velocity, 
and $\mathbf{I}^S = \mathbf{I} - \mathbf{n}\mathbf{n}$ is the surface identity tensor.
% $\nabla_s = \mathbf{I}^S \cdot \nabla$ denotes the surface gradient operator.

No-slip conditions are imposed at the boundaries. Together with continuity equation~\eqref{eq:contin}, we give at
$z=(-1)^j h_j$,
% and the second equality comes from equation~\eqref{eq:contin}. 
\begin{equation}
        \bm u_j=0\qquad \Rightarrow \quad \partial_zw_j=0.
        \label{eq:boundary_condition}
    \end{equation}

\subsection{Linear formulation and Fourier decomposition}

We linearise the governing equations about a quiescent base state with a flat interface. The base state is denoted by subscript $0$,  and is given by $\boldsymbol{u}_{j0}={0}$, $ p_j=p_{j0}(z)$, $ \Gamma=\Gamma_0$.
% where $p_{j0}$ is the hydrostatic pressure in fluid $j$. 
Infinitesimal perturbations are introduced as
\begin{equation}
z=\epsilon \zeta(\boldsymbol{x},t), \qquad
\boldsymbol{u}_j=\epsilon \boldsymbol{u}'_j, \qquad
p_j=p_{j0}(z)+\epsilon p'_j, \qquad
\Gamma=\Gamma_0+\epsilon \Gamma',
\label{eq:perturbation_expansion}
\end{equation}
with $\epsilon\ll 1$ and primed quantities of order unity.

Although the surface viscosities may depend on $\Gamma$, their concentration-dependent perturbations multiply the first-order surface strain rate and therefore enter only at $O(\epsilon^2)$. Hence only the base-state values $\eta_0^S$ and $\kappa_0^S$ enter the present linear analysis. $\eta_0^S=\eta^S(\Gamma_0)$ and $\kappa_0^S=\kappa^S(\Gamma_0)$ denote the base-state shear and dilatational surface viscosities, respectively.

Following \citet{kumar1994parametric},
we consider disturbances on an unbounded horizontal domain and assume normal-mode solutions of the form

\begin{equation}
w_j(\boldsymbol{x},z,t)=e^{i\boldsymbol{k}\cdot\boldsymbol{x}}w_j(z,t) \qquad \zeta(\boldsymbol{x},t)=e^{i\boldsymbol{k}\cdot\boldsymbol{x}}\zeta(t) \qquad 
\Gamma'(\boldsymbol{x},t)=e^{i\boldsymbol{k}\cdot\boldsymbol{x}}\Gamma'(t), 
\end{equation}
% $w(x,z,t)=e^{i\boldsymbol{k}x}w(z,t); \zeta(x,t)=e^{i\boldsymbol{k}x}\zeta(t);\Gamma'(x,z,t)=e^{i\boldsymbol{k}x}\Gamma'(z,t)$, 
where $\boldsymbol{k}=k_x\boldsymbol{e}_x+k_y\boldsymbol{e}_y$ is the horizontal wave vector and $k=|\boldsymbol{k}|$ is the 
% horizontal
wavenumber.

We now summarize the complete LSA problem. After eliminating the pressure from equation~\eqref{eq:NS}, the perturbation equations in the two fluid layers are obtained as

\begin{equation}
    (\partial_t-\nu_j(\partial_{zz}-k^2))(\partial_{zz}-k^2)w_j=0,
    \qquad
    -h_1 \leq z < 0 \ (j=1), \quad 0 < z \leq h_2 \ (j=2),
    \label{eq:eliminating horizontal v p linearized NS}
\end{equation}

\noindent where $w_j$ denotes the normal velocity perturbation and $\nu_j$ denotes the kinematic viscosity. The linearized surfactant transport equation at the interface is 

\begin{equation}
\label{eq:surfactant transport}
    \partial_t \Gamma' + D_s k^2 \Gamma' = \Gamma_0 \partial_z w|_{z=0}.
\end{equation}

The no-slip conditions at the two plates are given by
\begin{equation}
    w_1=\partial_z w_1=0 \quad (z=-h_1),
    \qquad
    w_2=\partial_z w_2=0 \quad (z=h_2),
    \label{eq:plates}
    \end{equation}
    
and the conditions at the interface are
\begin{equation}
    \jump{w}=0,\qquad \jump{\partial_z w}=0.
    \label{eq:int_continuity}
    \end{equation}
% The tangential stress balance yields
\begin{equation}
       \jump{\eta(\partial_{zz}+k^2)w} = \frac{\sigma_0\beta_s(k^2\Gamma')}{(1-\Gamma_0)} +(\kappa_0^S + \eta_0^S)k^2\partial_{z}w    
    \end{equation}

% The normal stress balance (after eliminating pressure term from linearized Navier-Stokes equation same method as \cite{kumar1994parametric}) reads
\begin{equation}
        \jump{(\rho\partial_t\partial_zw-\eta\partial_{zzz}w+3k^2\eta \partial_{z}w)}=k^4\sigma_b\zeta-k^2\jump{\rho}G(t)\zeta.
        \label{eq:int_normal_FHS}
    \end{equation}

The kinematic condition at the interface reads
\begin{align}
\partial_t \zeta &= w|_{z=0}. 
\label{kinematic_condition}
\end{align}

Equations~\eqref{eq:eliminating horizontal v p linearized NS}-\eqref{kinematic_condition} define 
the complete linearized
 hydrodynamic system, 
% governing the linear stability
for a surfactant-laden two-phase Faraday configuration with surface rheology.

\subsection{Floquet formulation}
We apply Floquet theory to analyze the linear stability of the time-periodic system. 
% Since the forcing $G(t)$ is periodic with period $T = 2\pi/\Omega$, 
% the linearized problem has time-periodic coefficients.
% % the governing equations have periodic coefficients in time.
% According to Floquet theory, solutions can be written as
% \begin{equation}
%     \bm q(z,t) = e^{\lambda t}\bm c(z,t), \qquad \bm c(z,t+T)=\bm c(z,t),
% \end{equation}
% where $\lambda$ is the Floquet exponent and $\mu = e^{\lambda T}$ is the Floquet multiplier and $\bm q$ denotes the collection of perturbation variables
% $(w_1,w_2,\zeta,\Gamma')$. The periodic function $\bm c(z,t)$ is expanded in a Fourier series,
% % \begin{equation}
% $\bm c(z,t)=\sum_{n=-N}^{N}\bm c_n(z)e^{\ii n\Omega t}$.
% % \end{equation}
%     % Thus 
% The solution takes the form
% \begin{equation}
%     \bm q(z,t)=\sum_{n=-N}^{N}\bm c_n(z)e^{s_n t}, 
%     \qquad 
%     s_n \equiv \lambda+\ii n\Omega.
% \end{equation}
% Here, $n$ denotes the temporal Floquet harmonic index. 
% % At neutral stability, $\mathrm{Re}(\lambda)=0$. 
% The harmonic and subharmonic neutral branches correspond to $\lambda=i\alpha\Omega$, with $\alpha=0$ and $\alpha=1/2$, respectively.
% % At neutral stability, one sets $\lambda=\ii\alpha\Omega$, with $\alpha=0$ (harmonic response) or $\alpha=\tfrac12$ (subharmonic response).
% % i.e., $ s_n = \ii (n+\alpha)\Omega$.
Because the forcing is periodic with $T=2\pi/\Omega$, we seek Floquet solutions
\begin{equation}
(w_j,\zeta,\Gamma')
=
\sum_{n=-N}^{N}
(w_{j,n},\zeta_n,\Gamma_n)e^{s_nt},
\qquad
s_n=\lambda+in\Omega.
\label{eq:floquet-expansion}
\end{equation}
At neutrality, $\lambda=i\alpha\Omega$, with $\alpha=0$ and $1/2$
for harmonic and subharmonic responses, respectively;   no 
% assumption is made on the 
phase relation between $\Gamma'$ and $\zeta$ is prescribed.
    
% Since all bulk equations and all interfacial conditions except the normal stress balance are time independent, they remain diagonal in the temporal harmonic index $n$
% after the Floquet expansion. Writing
% \begin{equation}
%     w_j(z,t)=\sum_{n=-N}^{N} w_{j,n}(z)\,e^{s_n t},\qquad
%     \zeta(t)=\sum_{n=-N}^{N} \zeta_n\,e^{s_n t},\qquad
%     \Gamma'(t)=\sum_{n=-N}^{N} \Gamma_n\,e^{s_n t},
% \label{eq:floquet-expansion}
% \end{equation}
% with    
% \begin{equation}
%     s_n \equiv \lambda + i n\Omega, \qquad 
%     \lambda=i\alpha\Omega \ \ (\alpha=0\ \text{harmonic},\ \alpha=\tfrac12\ \text{subharmonic}),
% \end{equation}
% where no assumption is made on the phase relation between
% $\Gamma'$ and $\zeta$, 
% the surfactant concentration and the interface deformation. 
% instead, this relation emerges naturally from the coupled Floquet eigenvalue problem. 
% In contrast to earlier approaches where the phase relation was prescribed \citep{kumar2002parametrically}, the present formulation determines this phase in the same idea as \citep{giavedoni2007onset}.

% We then obtain for each fixed $n$ the following decoupled boundary-value problem as follows.
Substituting \eqref{eq:floquet-expansion} into   \eqref{eq:eliminating horizontal v p linearized NS}, we obtain for each layer $j$ and 
each Fourier component  $n$,  a fourth-order ordinary differential equation in $z$
% Bulk equations (each layer)
\begin{equation}
    \label{eq:OS_n}
    \Big[s_n-\nu_j\big(\partial_{zz}-k^2\big)\Big]\big(\partial_{zz}-k^2\big)w_{j,n}=0,
    \qquad \nu_j=\eta_j/\rho_j,
\end{equation}
valid in $-h_1\le z\le 0$ for $j=1$ and $0\le z\le h_2$ for $j=2$.
The solution is given by

\begin{equation}
        w_{jn}(z)=b_{jn}e^{kz}+c_{jn}e^{-kz}+d_{jn}e^{q_{jn}z}+f_{jn}e^{-q_{jn}z}
        \label{eq:solution}
\end{equation}
where $q_{jn}^2=k^2+s_n/\nu_j$.\\
We enforce no-slip conditions at the plates
\begin{equation}
    \label{eq:plates_n}
    w_{1,n}(-h_1)=0,\qquad \partial_z w_{1,n}(-h_1)=0,
\end{equation}
\begin{equation}
    w_{2,n}(h_2)=0,\qquad \partial_z w_{2,n}(h_2)=0.
    \label{eq:plates2_n}
\end{equation}
The interfacial continuity conditions at $z=0$ are
\begin{equation}
    \label{eq:cont_w_n}
    \jump{w_n}\equiv w_{2,n}(0)-w_{1,n}(0)=0,
\end{equation}
\begin{equation}
    \label{eq:cont_Dw_n}
    \jump{\partial_z w_n}\equiv \partial_z w_{2,n}(0)-\partial_z w_{1,n}(0)=0.
\end{equation}
    Tangential stress balance with Marangoni forcing for each $n$,    
\begin{equation}
        \jump{\eta_j(\partial_{zz}+k^2)w_{n,j}}= \frac{\sigma_0\beta_s(k^2\Gamma_n)}{(1-\Gamma_0)} +k^2(\kappa_0^S + \eta_0^S)\partial_zw_n 
        \label{eq:ts}
\end{equation}
The kinematic condition $\partial_t\zeta=w|_{z=0}$ yields, for each $n$,
\begin{equation}
    \label{eq:kin_n}
    s_n \zeta_n = w_{1,n}(0)=w_{2,n}(0),
\end{equation}
where the equality follows from \eqref{eq:cont_w_n}.\\
The linearized  surfactant transport equation 
$\partial_t\Gamma'+D_s k^2\Gamma'=\Gamma_0\,\partial_z w|_{z=0}$ 
gives, for each temporal harmonic $n$,
\begin{equation}
    \label{eq:surf_n}
    \big(s_n + D_s k^2\big)\Gamma_n= \Gamma_0\,\partial_z w_{1,n}(0)=\Gamma_0\,\partial_z w_{2,n}(0).
\end{equation}

Then we can eliminate $\Gamma_n$ in equation \eqref{eq:ts} by using \eqref{eq:surf_n}
\begin{equation}
        \jump{\eta_j(\partial_{zz}+k^2)w_{n,j}} = \frac{\sigma_0\beta_s\Gamma_0}{(1-\Gamma_0)\big(s_n + D_s k^2\big)}k^2\partial_zw_n +(\kappa_0^S + \eta_0^S)k^2\partial_zw_n
        \label{eq:tangential_stress_balance}
\end{equation}   

% The time-periodic forcing enters through the vertical acceleration
% \begin{equation}
%     G(t)=g_0 + a\cos(\Omega t)=g_0+\frac{a}{2}(e^{\ii\Omega t}+e^{-\ii\Omega t}).
% \end{equation}

% Substitution of  the Floquet expansion into the normal stress condition \eqref{eq:int_normal_FHS} yields
Using $G(t)=g_0+a(e^{i\Omega t}+e^{-i\Omega t})/2$, the normal stress balance becomes
\begin{equation}
    \jump{(\rho s_n+3k^2\eta)\partial_zw_n-\eta\partial_{zzz}w_n}=k^4\sigma_b\zeta_n-k^2\jump{\rho}\left[g_0\zeta_n+\frac{a}{2}(\zeta_{n+1}+\zeta_{n-1})\right].
    \label{eq:normal_hill}
\end{equation}

Thus the vibration term couples $n$ to $n\pm1$, giving a block tridiagonal Floquet system, similar to the work by \cite{kumar1994parametric}.

\subsection{Generalized eigenvalue problem and validation }

After applying the boundary and interfacial conditions to the vertical basis functions and truncating the Floquet expansion to $|n|\le N$, the system can be written as
\begin{equation}
A\boldsymbol m=aB\boldsymbol m.
\end{equation}
Here, $A$ contains the bulk and time-independent interfacial contributions, while $B$ arises solely from the periodic acceleration and couples neighboring Floquet harmonics. The minimum neutral eigenvalue over $k$ determines the critical acceleration $a_c$ and wavenumber $k_c$.

The surfactant-free limit is validated against \citet{kumar1994parametric}, while the insoluble-surfactant and surface-rheology limits are validated against the free-surface results of \citet{ubal2005formation,ubal2005influence}.

\section{Stability characteristics 
}

We first examine separately the effects of Marangoni elasticity and surface viscosity. In the linearized problem, both mechanisms enter through the tangential stress balance,
\begin{equation}
\jump{\eta_j(k^2+\partial_{zz})w_{n,j}}=\underbrace{
\frac{\sigma_0\beta_s\Gamma_0}{(1-\Gamma_0)(s_n+D_sk^2)}\,k^2\partial_z w_n
}_{\text{Marangoni contribution}}+\underbrace{
(\kappa_0^S+\eta_0^S)k^2\partial_z w_n
}_{\text{surface-viscous contribution}}.
\label{eq:ts_ratio_base}
\end{equation}
Although both contributions act on the same interfacial strain-rate quantity, their temporal responses are fundamentally different. The surface-viscous coefficient is real, whereas the Marangoni contribution contains a complex surfactant-response factor $(s_n+D_sk^2)^{-1}$. For the critical subharmonic component, $s_0=i\Omega/2$, we therefore characterize the strengths of the two mechanisms by
\begin{equation}
C_M=\left|\frac{\sigma_0\beta_s\Gamma_0 k_c^2}{(1-\Gamma_0)(i\Omega/2+D_sk_c^2)}\right|,
\qquad
C_S=(\kappa_0^S+\eta_0^S)k_c^2.
\label{eq:CM_CS_def}
\end{equation}

\begin{figure}
\centering
\begin{tabular}{c}
\includegraphics[width=\textwidth]{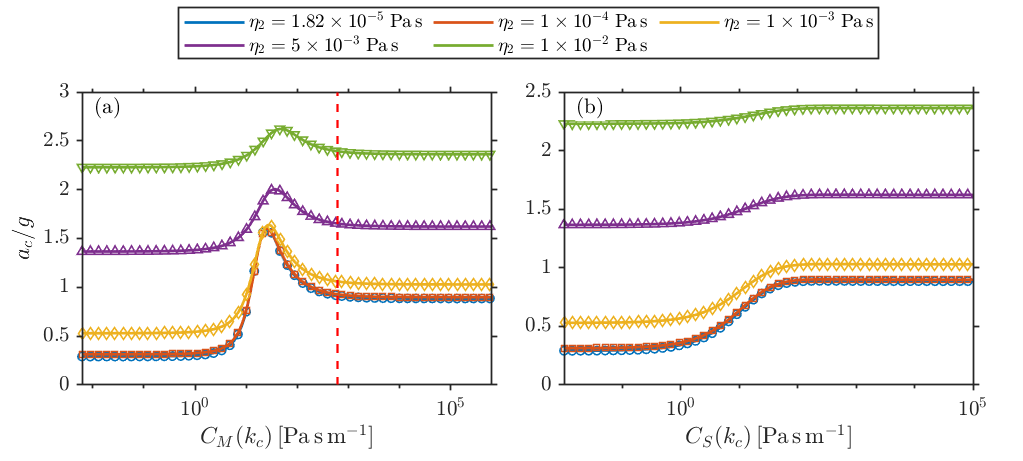}
\end{tabular}
\caption{Effect of (a)
% Marangoni elasticity 
$C_M(k_c)$ and (b) 
% surface viscosity 
$C_S(k_c)$ on 
% the critical acceleration
$a_c/g$ for different upper-layer viscosities in the
water–air reference density contrast.
% with parametrically varied upper-phase viscosity. 
The dashed red line marks the maximum admissible
% The dashed red line in (a) denotes the upper physically admissible value of
$C_M$ associated with the regularized Langmuir equation of state.}
\label{varying_CMCS}
\end{figure}

To examine these effects under representative two-fluid conditions, we consider a water--air configuration with $\rho_1=998~\mathrm{kg\,m^{-3}}$, $\rho_2=1.2~\mathrm{kg\,m^{-3}}$, $\eta_1=1.0\times10^{-3}~\mathrm{Pa\,s}$, $\sigma_0=0.06~\mathrm{N\,m^{-1}}$, and $\Omega=2\pi\times100~\mathrm{rad\,s^{-1}}$. The layer depths are $h_1=h_2=5l_c$, where $l_c=[\sigma_0/(|\rho_1-\rho_2|g)]^{1/2}$. The upper-layer viscosity $\eta_2$ is varied to examine the influence of the surrounding viscous fluid. A regularized Langmuir equation of state, $\sigma=\sigma_0(\max[0.05,1+\beta_s\ln(1-\Gamma)])$, is used, and the base concentration is restricted to $0\leq\Gamma_0\leq0.8$. Surface viscosities are varied over the range $10^{-8}$--$10^{-3}~\mathrm{Pa\,s\,m}$, consistent with experimentally reported values \citep{stevenson2005remarks,liu2006experimental}. In the Marangoni sweep, $\Gamma_0$, $D_s$, $\sigma_0$ and $\Omega$ are held fixed while $\beta_s$ is varied, and  $k_c$ is recomputed at each value of $\beta_s$; $C_M$ is then evaluated from \eqref{eq:CM_CS_def} using the corresponding $k_c$. For the surface-viscosity sweep, $\beta_s=0$ and $\kappa_0^S+\eta_0^S$ is varied, with $C_S$ likewise evaluated at
the corresponding $k_c$.

Figure~\ref{varying_CMCS} shows the dependence of the critical acceleration on $C_M$ and $C_S$ for different upper-layer viscosities. The two interfacial mechanisms are varied independently, with the other contribution set to zero. In both cases, the limits $C_M\rightarrow0$ and $C_S\rightarrow0$ recover the corresponding clean-interface threshold. 
Increasing $C_M$ or $C_S$ leads to a stabilizing effect as $a_c/g$ increases over the parameter range considered. 
However, for the Marangoni cases, the stabilization is strongly non-monotonic, while $C_S$ produces a monotonic increase in $a_c/g$.
For both cases, when $C_M\rightarrow \infty $ or $C_S \rightarrow \infty$, 
the threshold approaches a plateau at sufficiently large interfacial resistance.
% lead to a plateau, and reach a  similar critical accelerations for each upper-layer viscosity.
The origin of this common asymptote is discussed below. For $C_M$ and $C_S$,
the increase in critical acceleration becomes smaller as $\eta_2$ increases.
To isolate the Marangoni effect in the $C_M$ cases, we repeat the calculations at fixed equilibrium surface tension. The non-monotonic threshold persists for all upper-layer viscosities, demonstrating that it arises from Marangoni coupling rather than changes in the base-state surface tension.

To distinguish changes in modal damping from changes in resonance, we examine the corresponding unforced gravity--capillary modes by setting $a=0$.
% To understand the monotonic response to  $C_S$ and the non-monotonic response to $C_M$, we aim to separate changes in modal damping from changes in resonance. We therefore examine the corresponding unforced gravity--capillary eigenmodes by setting $a=0$.
For a prescribed wavenumber $k$, the  dominant interfacial  mode is characterized by a complex  temporal eigenvalue
% \begin{equation}
$\lambda_f=-\gamma_d+i\omega_0$,
% \label{eq:free_eigenvalue}
% \end{equation}
where $\gamma_d>0$ is the decay rate and $\omega_0$ is the natural  frequency. We characterize this mode using the normalized decay rate $\gamma_d/(\Omega/2)$, and the dimensionless detuning from the subharmonic resonance, 
% \begin{equation}
i.e., ${|\omega_0-\Omega/2|}/({\Omega/2})$.
The unforced problem obeys the same bulk equations and interfacial stress conditions as the forced system, but the temporal harmonics are no longer coupled by the periodic acceleration. For the Marangoni case,
 the surfactant response is  evaluated at the eigenvalue of the unforced mode.
% , giving the complex Marangoni coefficient
% \begin{equation}
% \widehat C_M(\lambda_f,k)=\frac{\sigma_0\beta_s\Gamma_0k^2}{(1-\Gamma_0)(\lambda_f+D_sk^2)}.
% \label{eq:free_CM}
% \end{equation}

\begin{figure}
\centering
\begin{tabular}{c}
\includegraphics[width=\textwidth]{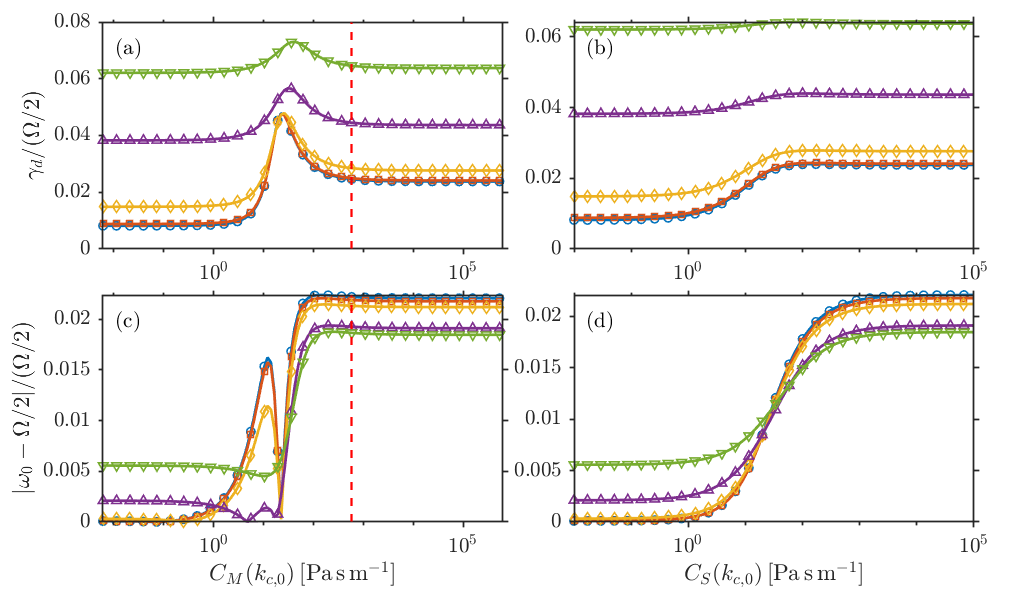}
\end{tabular}
\caption{Free-wave response at $k=k_{c,0}$: (a,b) normalized decay rate and (c,d) normalized detuning for the $C_M(k_{c,0})$ and $C_S(k_{c,0})$ sweeps, respectively. Colors denote the upper-layer viscosities identified in figure~\ref{varying_CMCS}.}
\label{fig:freewave_fixed}
\end{figure}

We first fix the wavenumber at the clean critical value, $k=k_{c,0}$,  to isolate the direct effect of interfacial rheology. Figure~\ref{fig:freewave_fixed} shows that both Marangoni  and surface viscosity modify the  decay rate and the natural frequency of the free interfacial mode.
Their effects on the damping, however, are qualitatively different. The decay rate $\gamma_d$ varies non-monotonically with $C_M$, whereas it increases monotonically with $C_S$; in both cases it approaches a limiting value at large interfacial resistance.  These trends are consistent with the corresponding non-monotonic and monotonic variations of the Faraday threshold.
At fixed $k_{c,0}$, both mechanisms also shift the free-mode frequency away from $\Omega/2$, producing normalized detunings of $O(10^{-2})$.  Thus, when the wavenumber is artificially constrained to remain at its clean-interface value, rheology modifies both the damping and the resonance condition.

\begin{figure}
\centering
\begin{tabular}{c}
\includegraphics[width=\textwidth]{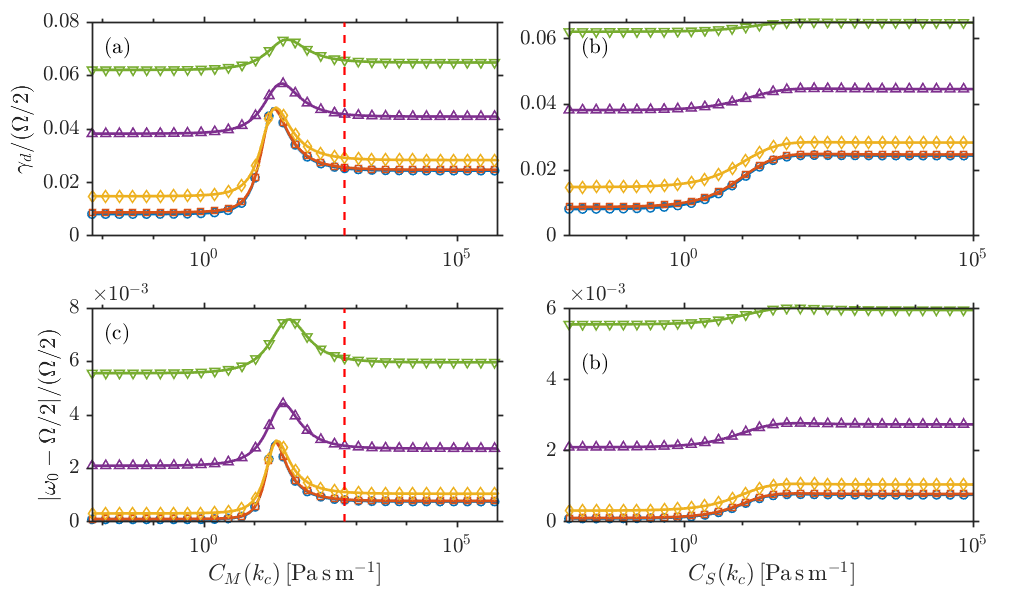}
\end{tabular}
\caption{Free-wave response along the Faraday critical branch $k=k_c$: (a,b) normalized decay rate and (c,d) normalized detuning for the $C_M(k_c)$ and $C_S(k_c)$ sweeps, respectively. Colors denote the upper-layer viscosities identified in figure~\ref{varying_CMCS}.}
\label{fig:freewave_path}
\end{figure}

The critical Faraday mode, however, is free to adjust its wavenumber. 
We repeat the free-wave calculation along $k=k_c(C_M)$ and $k=k_c(C_S)$. 
Figure~\ref{fig:freewave_path} shows  that the detuning is then reduced to 
% Along these branches, the detuning is reduced to 
$O(10^{-3})$. The shift in the critical wavenumber therefore largely compensates the rheology-induced change in the natural frequency and keeps the selected mode close to subharmonic resonance. 
Along the selected branches, the decay rate follows the same qualitative trends as the instability threshold: both $a_c/g$ and $\gamma_d$ vary non-monotonically with $C_M$, whereas both increase monotonically and approach a plateau with increasing $C_S$. Moreover  $|\omega_0-\Omega/2|/\gamma_d \lesssim 0.1$, so that the residual detuning is small compared with the modal damping. 
To assess the correspondence quantitatively, we compare the normalized critical acceleration and decay rate through $(a_c/a_{c,0})$/$(\gamma_d/\gamma_{d,0})$. This ratio remains close to unity,  with deviations below $2\%$ for surface viscosity and up to $7\%$
at intermediate Marangoni strength.
These results indicate that changes in free-wave damping largely account for the threshold variation, while residual detuning plays a secondary role.

The different damping responses are consistent with the different forms of the tangential interfacial stress. Surface viscosity provides a direct dissipative resistance, whereas the Marangoni stress depends on the evolving surfactant field. For the unforced mode, this response contains the complex factor $(\lambda_f+D_s k^2)^{-1}$, so that changes in Marangoni elasticity modify both the magnitude and the phase of the tangential stress relative to the interfacial motion. This provides a possible mechanism for the non-monotonic variation of the modal damping with $C_M$. By contrast, the surface-viscous contribution acts directly through the interfacial strain rate, consistent with the monotonic increase and eventual saturation of $\gamma_d$ with $C_S$.

\subsection{Interfacial response in the strong-rheology limit}

\begin{figure}
\centering
\begin{tabular}{c}
\includegraphics[width=\textwidth]{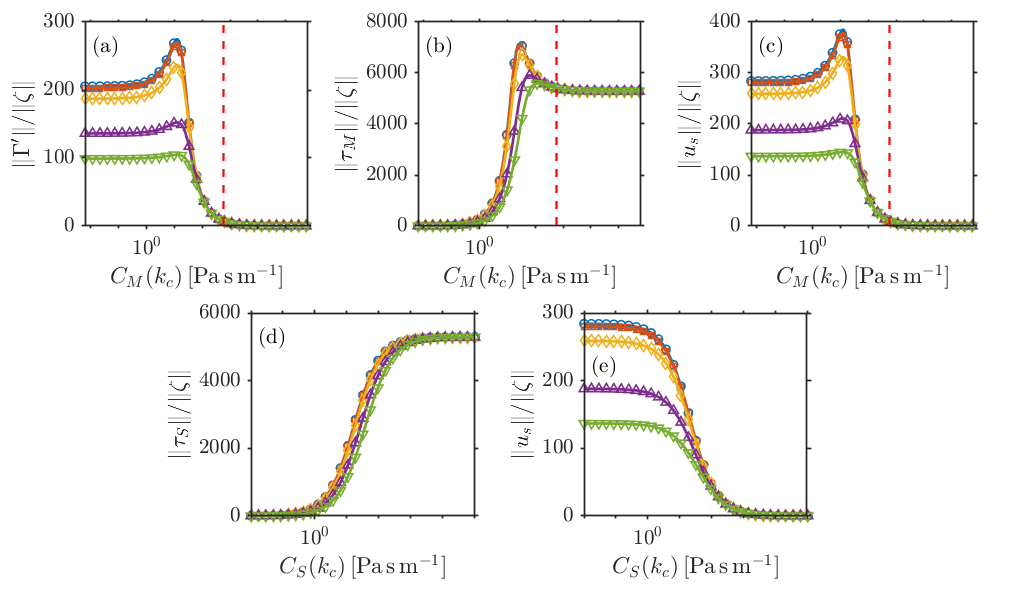}
\end{tabular}
\caption{Interfacial response of the critical Floquet eigenvector along the Marangoni and surface-viscosity branches. Panels (a--c) show $\Gamma'$, $\tau_M$ and $u_s$ versus $C_M(k_c)$, respectively, while panels (d,e) show $\tau_S$ and $u_s$ versus $C_S(k_c)$. All quantities are normalized by $\|\zeta\|_{\mathrm{RMS}}=1$. Colours denote the upper-layer viscosities identified in figure~\ref{varying_CMCS}.}
\label{fig:CM_eigenvector}
\end{figure}

The strong-rheology limit can be examined directly from the critical Floquet eigenvectors. Figure~\ref{fig:CM_eigenvector} shows $\Gamma'$, $\tau_M$ and $u_s$ for the Marangoni sweep in panels (a--c), and $\tau_S$ and $u_s$ for the surface-viscosity sweep in panels (d,e). Each eigenvector is normalized such that $\|\zeta\|_{\mathrm{RMS}}=1$. As $C_M$ increases, both $\|\Gamma'\|$ and $\|u_s\|$ vanish while $\|\tau_M\|$ remains finite; similarly, increasing $C_S$ suppresses $\|u_s\|$ while $\|\tau_S\|$ approaches a finite plateau.

For a Fourier mode with horizontal wavevector $\boldsymbol{k}$, the component of the interfacial velocity parallel to $\boldsymbol{k}$ is defined as $u_{s,n}=\frac{\boldsymbol{k}}{k}\cdot\boldsymbol{u}_{s,n}.$
The incompressibility condition then gives $ik\,u_{s,n}+\partial_z w_n(0)=0.$
The surfactant-transport condition, $(s_n+D_sk^2)\Gamma_n=\Gamma_0\partial_z w_n(0),$
then implies $\Gamma_n\rightarrow0$ as the interface becomes tangentially immobile. The 
$C_M \rightarrow \infty$ limit
% strong-Marangoni limit therefore
corresponds to the interface approaching the uniform base concentration $\Gamma_0$.

The persistence of a finite Marangoni stress in the strong-elasticity limit follows directly from the tangential stress condition. Eliminating $\Gamma_n$ gives
\begin{equation}
\jump{\eta_j(\partial_{zz}+k^2)w_{n,j}}
=
\left[
\frac{\sigma_0\beta_s\Gamma_0k^2}
{(1-\Gamma_0)(s_n+D_sk^2)}
+
(\kappa_0^S+\eta_0^S)k^2
\right]
\partial_z w_n(0).
\label{eq:strong_tangential}
\end{equation}
For a pure Marangoni sweep, define
\begin{equation}
\widehat C_{M,n}
=
\frac{\sigma_0\beta_s\Gamma_0k^2}
{(1-\Gamma_0)(s_n+D_sk^2)} .
\end{equation}
For a bounded, regularly normalized critical mode,
the left-hand side of \eqref{eq:strong_tangential} remains $O(1)$ as
$|\widehat C_{M,n}|\rightarrow\infty$. It follows that
\[
\partial_z w_n(0)
=O(|\widehat C_{M,n}|^{-1}).
\]
Since incompressibility gives
$u_{s,n}\propto\partial_z w_n(0)$, the tangential interfacial velocity therefore vanishes as the Marangoni resistance increases. The surfactant transport equation correspondingly gives $\Gamma_n=O(|\widehat C_{M,n}|^{-1})$, while the product producing the Marangoni stress remains $O(1)$. Thus the finite Marangoni stress acts as a reaction stress enforcing tangential interfacial immobilization. The same argument applies to a pure surface-viscosity sweep. In this case the interfacial contribution is proportional to $C_S\partial_z w_n(0)$, so bounded bulk stresses require $\partial_z w_n(0)\sim C_S^{-1}$ and hence $u_s\sim C_S^{-1}$, while the surface-viscous stress remains $O(1)$.

These asymptotic scalings are confirmed by log--log fits over the largest decade of $C_M$ and $C_S$. For all upper-layer viscosities considered, the fitted exponents are approximately $-1$ for both $\|\Gamma'\|$ and $\|u_s\|$ in the Marangoni sweep, whereas $\|\tau_M\|$ approaches a constant. Similarly, in the surface-viscosity sweep, $\|u_s\|\propto C_S^{-1}$ while $\|\tau_S\|$ approaches a constant. Hence both mechanisms approach the same tangentially
immobile interfacial condition. When the equilibrium surface tension and all remaining parameters are held fixed, the corresponding Floquet problems therefore converge to the same asymptotic onset threshold.

\section{Conclusions}

We have extended the two-fluid Floquet formulation for Faraday instability to include insoluble-surfactant transport, Marangoni stresses and Boussinesq–Scriven surface viscosity. Over the parameter range considered, both interfacial mechanisms increase the onset acceleration relative to the clean-interface state, but their finite-strength responses are qualitatively different: the threshold varies non-monotonically with Marangoni stresses, and monotonically towards saturation with surface viscosity.

Analysis of the corresponding unforced gravity–capillary mode separates changes in damping from changes in resonance. When the wavenumber is fixed at its clean-interface value, both effects are appreciable. Along the selected Faraday critical branch, however, wavenumber reselection strongly reduces the detuning from subharmonic resonance.
% and the variation of the onset threshold closely follows that of the selected-mode decay rate. 
The critical acceleration largely follows the decay rate of the selected mode. Thus, modal damping provides the leading explanation for the threshold variation, while residual detuning plays a secondary role.

The two mechanisms also share a common formal strong-rheology limit. Increasing Marangoni elasticity drives the surfactant perturbation and tangential velocity towards zero while maintaining a finite reaction stress; increasing surface viscosity suppresses the same tangential motion directly through viscous resistance. For fixed equilibrium surface tension, both therefore approach the same tangentially immobile-interface Floquet problem. These results distinguish the effects of Marangoni elasticity and surface viscosity at finite strength while identifying their common limiting behaviour.\\

\noindent \textbf{Declaration of Interests}. The authors report no conflict of interest. \\

\noindent\textbf{Data availability.} 
The code  supporting the findings of this study is publicly available 
at 
\url{https://github.com/CFTL-Illinois/Faraday_surface_rheology_LSA}.
% \\

\bibliographystyle{jfm}
\bibliography{MAIN.bib}

@article{kumar1994parametric,
  title={Parametric instability of the interface between two fluids},
  author={Kumar, K. and Tuckerman, L. S.},
  journal={J. Fluid Mech.},
  volume={279},
  pages={49--68},
  year={1994}
}

@article{benjamin1954stability,
  title={The stability of the plane free surface of a liquid in vertical periodic motion},
  author={Benjamin, T. B. and Ursell, F. J.},
  journal={Proc. R. Soc. A},
  volume={225},
  number={1163},
  pages={505--515},
  year={1954}
}

@article{faraday1831xvii,
  title={On a peculiar class of acoustical figures and on certain forms assumed by groups of particles upon vibrating elastic surfaces},
  author={Faraday, M.},
  journal={Phil. Trans. R. Soc. Lond.},
  volume={121},
  pages={299-340},
  year={1831}
}

@article{HENDERSON_1998, title={Effects of surfactants on Faraday-wave dynamics}, volume={365}, DOI={10.1017/S0022112098001086}, journal={J. Fluid Mech.}, author={Henderson, D. M.}, year={1998}, pages={89–107}}

@article{miles1990parametrically,
  title={Parametrically forced surface waves},
  author={Miles, J. and Henderson, D.},
  journal={Annu. Rev. Fluid Mech.},
  volume={22},
  number={1},
  pages={143--165},
  year={1990}
}

@article{kumar2002parametrically,
  title={Parametrically driven surface waves in surfactant-covered liquids},
  author={Kumar, S. and Matar, O. K.},
  journal={Proc. R. Soc. A},
  volume={458},
  number={2027},
  pages={2815--2828},
  year={2002}
}

@article{kumar2004faraday,
  title={On the {F}araday instability in a surfactant-covered liquid},
  author={Kumar, S. and Matar, O. K.},
  journal={Phys. Fluids},
  volume={16},
  number={1},
  pages={39--46},
  year={2004}
}

@article{ubal2005elastic,
  title={Elastic effects of an insoluble surfactant on the onset of two-dimensional {F}araday waves: a numerical experiment},
  author={Ubal, S. and Giavedoni, M. D. and Saita, F. A.},
  journal={J. Fluid Mech.},
  volume={524},
  pages={305--329},
  year={2005}
}

@article{ubal2005formation,
  title={The formation of {F}araday waves on a liquid covered with an insoluble surfactant: influence of the surface equation of state},
  author={Ubal, S. and Giavedoni, M. D. and Saita, F. A.},
  journal={Lat. Am. Appl. Res.},
  volume={35},
  number={1},
  pages={59--66},
  year={2005}
}

@article{ubal2005influence,
  title={Influence of surface viscosity on two-dimensional {F}araday waves},
  author={Ubal, S. and Giavedoni, M. D. and Saita, F. A.},
  journal={Ind. Eng. Chem. Res.},
  volume={44},
  number={4},
  pages={1090--1099},
  year={2005}
}

@article{stevenson2005remarks,
  title={Remarks on the shear viscosity of surfaces stabilised with soluble surfactants},
  author={Stevenson, P.},
  journal={J. Colloid Interface Sci.},
  volume={290},
  number={2},
  pages={603--606},
  year={2005}
}

@article{liu2006experimental,
  title={An experimental study of surfactant effects on spilling breakers},
  author={Liu, X. and Duncan, J. H.},
  journal={J. Fluid Mech.},
  volume={567},
  pages={433--455},
  year={2006}
}

@article{giavedoni2007onset,
  title={Onset of {F}araday waves in a liquid layer covered with a surfactant with elastic and viscous properties},
  author={Giavedoni, M. D. and Ubal, S.},
  journal={Ind. Eng. Chem. Res.},
  volume={46},
  number={15},
  pages={5228--5237},
  year={2007}
}

@book{brenner2013interfacial,
  title={Interfacial Transport Processes and Rheology},
  author={Brenner, H.},
  publisher={Elsevier},
  year={2013}
}

@article{scriven1960dynamics,
  title={Dynamics of a fluid interface equation of motion for Newtonian surface fluids},
  author={Scriven, L. E.},
  journal={Chem. Eng. Sci.},
  volume={12},
  number={2},
  pages={98--108},
  year={1960}
}

@article{bird2002transport,
  title={Transport phenomena},
  author={Bird, R. B.},
  journal={Appl. Mech. Rev.},
  volume={55},
  number={1},
  pages={R1--R4},
  year={2002}
}

@article{strickland2024role,
  title={The role of higher-order viscous and interfacial effects on the onset of surfactant-covered faraday waves},
  author={Strickland, Stephen L and Daniels, Karen E and Shearer, Michael},
  journal={IMA Journal of Applied Mathematics},
  volume={89},
  number={6},
  pages={1006--1040},
  year={2024},
  publisher={Oxford University Press}
}

\end{document}